\documentclass[aps,prl,reprint,groupedaddress,nofootinbib,nopreprintnumbers]{revtex4-2}

\usepackage{amsfonts}
\usepackage{amsmath}
\usepackage{amssymb}
\usepackage{amsthm}
\usepackage{mathtools}
\usepackage{braket}
\usepackage{stackrel}
\usepackage{tensor}
\usepackage{cancel}
\usepackage{enumerate}
\usepackage{esint}
\usepackage{graphicx}
\usepackage{subcaption}
\usepackage{hyperref}
\usepackage{color}
\usepackage[capitalise]{cleveref}

\newcommand{\sgn}{\text{sgn}}
\newcommand{\ammaG}{\text{\reflectbox{$\Gamma$}}}
\renewcommand{\Re}{\text{Re}}
\renewcommand{\Im}{\text{Im}}

\newcommand{\eg}{\textit{e.g.,} }

\begin{document}

\title{The negativity core of a 1+1D massless scalar quantum field}

\author{Jason Pye}
\email{jason.pye@su.se}
\affiliation{Nordita, Stockholm University and KTH Royal Institute of Technology, Hannes Alfvéns väg 12, SE-106 91 Stockholm, Sweden}

\author{Atharva Hingane}
\email{atharvah@iisc.ac.in}
\affiliation{Department of Computational and Data Sciences, Indian Institute of Science, Bangalore 560012, India}

\author{Robert H.~Jonsson}
\email{robert.jonsson@mau.se}
\affiliation{Department of Materials Science and Applied Mathematics, Malmö University, SE-205 06, Malmö, Sweden}
\affiliation{Nordita, Stockholm University and KTH Royal Institute of Technology, Hannes Alfvéns väg 12, SE-106 91 Stockholm, Sweden}

\date{22 May 2026}

\preprint{NORDITA-2026-063}

\begin{abstract}
    Vacuum entanglement is a fundamental feature of quantum field theory exhibiting rich structure that is not completely understood.
    Here, we provide a complete characterization of the entanglement between two bounded spacelike-separated regions in a (1+1)-dimensional free massless real scalar field.
    Employing Gaussian state methods, we analytically compute the logarithmic negativity and construct closed-form solutions for the localized modes carrying it, called negativity cores.
    These results deepen our understanding of quantum fields and suggest extensions to higher dimensions and fermionic fields.
\end{abstract}

\maketitle

The entanglement structure of quantum fields plays a central role in various research areas, from providing tools in studying the properties of quantum many-body systems~\cite{amico_entanglement_2008,eisert_area_2010} and in modern developments of holography~\cite{ryu_holographic_2006,swingle_entanglement_2012,almheiri_entropy_2019,penington_entanglement_2020}, to the possibility that spacetime itself may emerge from entanglement~\cite{vanraamsdonk_building_2010,jacobson_entanglement_2016,cao_space_2017}.
These ideas are rooted in foundational works on the connection of entanglement entropy and the geometry of local subsystems in quantum field theory (QFT), particularly in the context of the area law for black hole entropy~\cite{sorkin_entropy_1983,bombelli_quantum_1986,srednicki_entropy_1993,callan_geometric_1994,solodukhin_entanglement_2011}.
At a more fundamental level, the Reeh-Schlieder theorem demonstrates that entanglement is intrinsic to the structure of relativistic QFT~\cite{redhead_ado_1995,witten_aps_2018}, with consequences such as obstructions to particle localizability~\cite{halvorson_particles_2002,papageorgiou_impact_2019}.

In this work, we address key questions concerning the entanglement between bounded spacelike-separated regions of a field, a physical scenario of fundamental interest.
Early work, in the framework of algebraic QFT, demonstrated violations of Bell inequalities and nonzero distillable entanglement between spacelike-separated regions~\cite{summers_maximal_1987,verch_distillability_2005}. 
It was then demonstrated that this entanglement can be extracted into causally-separated localized probes interacting with the field~\cite{valentiniNonlocalCorrelationsQuantum1991,reznikEntanglementVacuum2003,steegEntanglingPowerExpanding2009}---a phenomenon called entanglement harvesting---widely studied in relativistic quantum information (see~\cite{martin-martinezEntanglementCurvedSpacetimes2014,caribeLensingVacuumEntanglement2023} and references therein).

Entanglement across a bipartition of a pure state is typically quantified by the entanglement entropy, which has been well-studied in QFT by various methods~\cite{calabrese_entanglement_2004,calabrese_entanglement_2009,casini_entanglement_2009,casini_lectures_2022,bianchiEntropySubalgebraObservables2019}.
The reduced state of a field in two bounded spacelike-separated regions, however, is mixed.
In this case, a commonly-used measure is the logarithmic negativity, which quantifies the extent to which a state violates the positive partial transpose (PPT) criterion, and is an upper bound to the distillable entanglement~\cite{horodecki_quantum_2009,serafini_quantum_2023}.
It has been computed numerically for lattice models~\cite{marcovitch_critical_2009,klco_entanglement_2021}, and its asymptotic scaling for close separation of regions in a 1+1D conformal field theory is known~\cite{calabrese_entanglement_2012,calabrese_entanglement_2013}.
Recently, an analytical calculation of the logarithmic negativity using replica methods was presented~\cite{arias_entanglement_2026}.

Gaussian state methods (see, \eg~\cite{serafini_quantum_2023}) both facilitate the calculation of entanglement measures and elucidate the structure of entanglement among the field degrees of freedom (modes).
We here tackle the question of which field modes in a region are most strongly entangled with those in another.
This has been studied numerically in lattice models and with particular families of smearing functions and detector models~\cite{zych_entanglement_2010,klco_entanglement_2023,gao_partial_2024,gao_detecting_2025,agullo_multimode_2025}.
For pure Gaussian states, it is well-known that a subsystem and its complement factor into a set of entangled pairs called partner modes~\cite{boteroModeWiseEntanglementGaussian2003,boteroSpatialStructuresLocalization2004a,hackl_minimal_2019,agulloCorrelationEntanglementPartners2025}.
A series of recent works~\cite{klco_entanglement_2023,gao_partial_2024,gao_detecting_2025} discovered that for certain classes of Gaussian states (including the vacuum of a free scalar field), the mixed state entanglement between two subsystems decomposes similarly into two-mode pairs, called negativity cores.
This decomposition therefore yields the optimal detection profiles for the entanglement between the two regions, which were computed numerically in a lattice model~\cite{gao_detecting_2025}.

In this paper, we fully characterize the logarithmic negativity between two arbitrary disjoint intervals in 1+1 dimensions for the vacuum state of a free massless real scalar field.
In particular, we provide (1) an analytical calculation of the logarithmic negativity using phase space methods, (2) the modewise decomposition of the subsystems in terms of the negativity cores, and (3) closed form solutions for the negativity core mode functions.\footnote{Our findings thus support the replica method results of~\cite{arias_entanglement_2026} for the total logarithmic negativity (which we became aware of after completing our findings) and, beyond that, yield a complete picture of the pairwise entanglement structure.}
To obtain our results, we use a framework based on the K\"ahler structure of Gaussian states~\cite{hackl_bosonic_2021}.
We reduce the problem to the diagonalization of an integral operator (closely related to the negativity Hamiltonian~\cite{murciano_negativity_2022}), which we show can be achieved by modifying a method of~\cite{arias_entropy_2018}.
The operator's eigenvalues determine the logarithmic negativity, and its eigenfunctions are used to construct the negativity cores following~\cite{klco_entanglement_2023,gao_partial_2024,gao_detecting_2025}.
We numerically validate our analytical results in a lattice model, and conclude by discussing possible extensions and future research directions.
This paper presents our main results, with complete technical details presented in an accompanying paper \cite{longer_paper}.

\begin{figure*}[t]
\centering
\includegraphics[width=\linewidth]{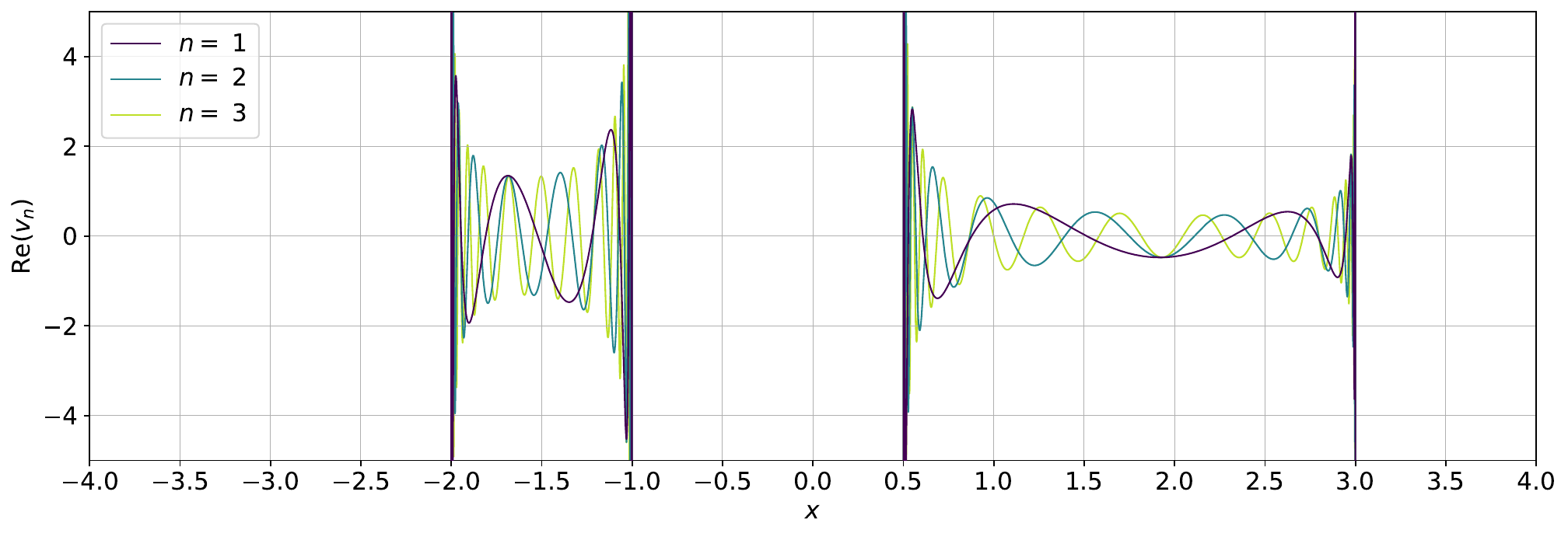}
\caption{Smearing functions defining the negativity cores for the three modes with the largest contributions to the negativity for $A = (-2,-1)$ and $B = (0.5,3)$.}
\label{fig:mode_profiles}
\end{figure*}

\textbf{\textit{Summary of results.---}}We consider the vacuum state of a $(1+1)$-dimensional free massless real scalar field.
This system is of fundamental interest, but it also models, \eg the electromagnetic vacuum in a transmission line.
Its Hamiltonian is
\begin{align}\label{eq:Hamiltonian}
    H = \frac12 \int dx \, [ \pi(x)^2 + \phi'(x)^2 ],
\end{align}
with canonical commutation relations $[ \phi(x), \pi(x') ] = i \delta(x-x')$.
The right- and left-moving modes in this model are decoupled, hence we decompose $\phi(x) = \phi_R(x) + \phi_L(x)$ and $\pi(x) = \pi_R(x) + \pi_L(x)= -\phi_R'(x) + \phi_L'(x)$.
These fields obey the commutation relations $[ \phi_R(x), \phi_R(x') ] = - [ \phi_L(x), \phi_L(x') ] = \tfrac{i}{4} \sgn(x-x')$.
The vacuum state $\ket{0}$ of the field is a Gaussian state, thus it is fully characterized by its one- and two-point functions.
The one-point function vanishes, $\bra{0} \phi_{R/L}(x) \ket{0} = 0$, while the two-point functions are
\begin{align}\label{eq:GRL}
  G_{R/L}(x,x') :=& \bra{0} \{ \phi_{R/L}(x), \phi_{R/L}(x') \} \ket{0} \nonumber\\
  =& \int \frac{dk}{2\pi} \frac{1}{2|k|} e^{i k (x-x')},
\end{align}
and $\bra{0} \{ \phi_R(x), \phi_L(x') \} \ket{0} = 0$.

The subsystems we consider correspond to two spacelike-separated observers with access to the field in the intervals $A = (a_1,a_2)$ and $B = (b_1,b_2)$, respectively.
We assume that the observers only couple to the derivatives of the field and not directly to its amplitude.
This is physically motivated by the fact that the amplitude does not enter the Hamiltonian density, and it is mathematically implemented by the condition that the smearing functions used to define linear observables $\int dx \, f_{R/L}(x) \phi_{R/L}(x)$ have a vanishing integral,
\begin{align}
  \int_A dx \, f_{R/L}(x) = 0 \quad \text{and} \quad \int_B dx \, f_{R/L}(x) = 0.
\label{eq:vanishing_integral}
\end{align}
This condition is also necessary to avoid coupling to the zero modes of $\phi_{R/L}(x)$, which cause divergences of the correlation functions.

Since the right- and left-moving fields are decoupled, they contribute equally to the logarithmic negativity.
We find that the logarithmic negativity of each of these sectors is given by $E_\mathcal{N} = - \sum_{n \in \mathbb{N}} \log_2 ( \tanh (\pi s_n) )$, where $\{ s_n \}_{n \in \mathbb{N}}$ are the discrete set of solutions to
\begin{align}
P_{-is-\tfrac12} \left( \frac{1+\eta}{1-\eta} \right) = 0, \label{eq:conical_zeros}
\end{align}
where $P_{-is-1/2}$ is a conical (Mehler) function~\cite{zhurina_tables_1966,hobson_theory_1931}, 
and where $\eta$ is the cross ratio
\begin{align}
    \eta = \frac{(a_2-a_1)(b_2-b_1)}{(b_1-a_1)(b_2-a_2)}.
\end{align}
Note that the cross ratio simplifies to $\eta = 1/(1+d/\ell)^2$, if the intervals are both of size $\ell=a_2-a_1=b_2-b_1$ and separated by $d=b_1-a_2$.

Each $s_n$ corresponds to a pair of modes defining a negativity core.
The core modes are identified by their 
quadratures in $A$,
\begin{align}
  \hat{Q}_{nA} &= \int_{a_1}^{a_2} dx \, \sqrt{2} \, \Re(v_{nA}(x)) \, \hat{\phi}_R(x), \nonumber\\
  \hat{P}_{nA} &= \int_{a_1}^{a_2} dx \, \sqrt{2} \, \Im(v_{nA}(x)) \, \hat{\phi}_R(x),\label{eq:quadratures}
\end{align}
and similarly for $B$, with the smearing functions (see \cref{fig:mode_profiles}) defined by
\begin{align}
  v_{nA}(x) = \frac{ \sqrt{2} C_n e^{-i s_n \omega(x)} }{ \sqrt{-(x-a_1)(x-a_2)(x-b_1)(x-b_2)} }, \label{eq:coremode_A}
\end{align}
for $x \in A = (a_1,a_2)$, and
\begin{align}
  v_{nB}(x) = \frac{ -\sqrt{2} C_n^\ast e^{i s_n \omega(x)} }{ \sqrt{-(x-a_1)(x-a_2)(x-b_1)(x-b_2)} }, \label{eq:coremode_B}
\end{align}
for $x \in B = (b_1,b_2)$, where $C_n$ is a normalization constant and $\omega(x) := \ln [ - (x-a_1)(x-b_2) / (x-a_2)(x-b_1) ]$.

The negativity simplifies in the two limiting regimes of large and small separation. 
In the regime where the intervals are widely separated, $d/\ell \gg 1$, the logarithmic negativity between each pair of modes is exponentially decaying as $E_{\mathcal{N},n} \sim 2 e^{-\pi j_{0,n} \left( 1 + \frac{d}{\ell} \right)}$, where $j_{0,n}$ is the $n^{th}$ zero of the Bessel function $J_0$~\cite{zhurina_tables_1966}.
The total logarithmic negativity is dominated by the smallest eigenvalue $n=1$, for which the decay rate is $\pi j_{0,1} \approx 7.555$.
In the regime where the separation goes to zero, we find that the total logarithmic negativity diverges as
\begin{align}\label{eq:lognegzerosep}
  E_\mathcal{N} \sim -\frac18 \log_2(1-\eta) - \log_2(-\ln(1-\eta)).
\end{align}
This leading order divergence agrees with the universal 1+1D conformal field theory result in~\cite{calabrese_entanglement_2012,calabrese_entanglement_2013} (after including a factor of 2 to account for the contributions from left-moving modes).

We will now describe the methods used to obtain our results.

\textbf{\textit{Diagonalization of $J^\ammaG$.---}}The negativity between two subsystems $A$ and $B$ is calculated from the spectrum of the partially transposed density operator of the system.
For Gaussian states, this translates into the spectrum of its partially transposed linear complex structure $J^\ammaG$~\cite{hackl_bosonic_2021,longer_paper}.
Specifically, the logarithmic negativity is given by $E_\mathcal{N} = \sum_n \text{max} \{ 0, -\log_2 \tilde{\nu}_n \}$, where $\{ \pm i \tilde{\nu}_n \}$ are the eigenvalues of the operator
\begin{align}
  J^\ammaG = \omega G^\Gamma.
\end{align}
The inverse symplectic form, $\omega = \omega_R \oplus \omega_L$, is represented by the kernels $\omega_R(x,x') = -\omega_L(x,x') = 2 \tfrac{d}{dx} \delta(x-x')$.
The bilinear form $G^\Gamma$ is constructed by first restricting $G = G_R \oplus G_L$ (with $G_{R/L}$ given in \cref{eq:GRL}) to the subspace $V_A^\ast \oplus V_B^\ast$ (where $V_{A/B}^\ast$ is the subspace of smearing functions supported in $A/B$ satisfying \cref{eq:vanishing_integral}), and then applying partial transposition, $G^\Gamma := \Gamma G \Gamma^T$.
In~\cite{longer_paper}, we show that $\Gamma^T$ can be implemented by a reflection in $B$, $x \mapsto b_1 + b_2 - x$ for $x \in (b_1,b_2)$, acting separately on the left- and right-moving sectors.
Since $\phi_{R/L}(x)$ are decoupled, we find $J^\ammaG = J_R^\ammaG \oplus J_L^\ammaG$ with
\begin{align}
  J_R^\ammaG(x,x') = \begin{cases}
  -\frac{1}{\pi} \text{P.V.} \frac{1}{x-x'}, &
  \substack{
    \text{if } x,x' \in A 
    \text{ or } x,x' \in B
  } \\
  -\frac{1}{\pi} \text{P.V.} \frac{1}{x+x'-b_1-b_2}, &
  \substack{
    \text{if } x \in A, x' \in B \\
    \text{ or } x \in B, x' \in A
  }
\end{cases}
\end{align}
and $J_L^\ammaG = -J_R^\ammaG$.
The eigenvalues of $J^\ammaG$ come in pairs $\pm i \nu_n$.
Due to the block structure of $J^\ammaG$, the right- and left-moving fields contribute equally to the total logarithmic negativity, thus below we will focus only on $J_R^\ammaG$ and drop the subscript $R$ for simplicity.

We explicitly compute the eigenvalues and eigenfunctions of $J^\ammaG$ by modifying a method introduced in~\cite{arias_entropy_2018} for the operator $(\omega G)(x,x') = - \tfrac{1}{\pi} \text{P.V.} \frac{1}{x-x'}$ on $A \cup B$.
We adapt this method to account for the partial transposition of $G$.
It reformulates the eigenvalue problem as a boundary value problem in the complex plane.
Let $A \cup B$ be two segments of the real axis in $\mathbb{C}$.
For some $0 \neq \lambda \in \mathbb{C}$, we construct a function $S(z)$ satisfying:
\begin{enumerate}[i)]
  \item $S(z)$ is analytic on $\mathbb{C} \setminus \overline{A \cup B}$, \label{cond:BVP_1}
  
  \item $S^+(x) = \begin{cases}
    \lambda S^-(x), & x \in A, \\ \lambda^{-1} S^-(x), & x \in B,
  \end{cases}$ \label{cond:BVP_2}
  
  \item $\lim_{|z| \to \infty} |z^2 S(z)| < \infty$, \label{cond:BVP_3}

  \item $\lim_{z \to p} | (z-p) S(z) | = 0$, where $p \in \{ a_1, a_2, b_1, b_2 \}$, \label{cond:BVP_4}
\end{enumerate}
where $S^\pm(x) := S(x+i0^\pm)$.
Solutions to this problem are in one-to-one correspondence with formal eigenfunctions of $J^\ammaG$, with eigenvalue $i\tilde{\nu} = i(1+\lambda^{-1})/(1-\lambda^{-1})$.
The valid eigenfunctions of $J^\ammaG$ are given by the boundary values $S^+(x)$ for $x \in A$ and $-S^-(b_1+b_2-x)$ for $x \in B$ which also satisfy the vanishing integral conditions of $V_A^\ast \oplus V_B^\ast$, i.e., $\int_{a_1}^{a_2} dx \, S^+(x) = \int_{b_1}^{b_2} dx \, S^-(x) = 0$.

Here we are only interested in eigenvalues contributing to the logarithmic negativity, $|\tilde{\nu}| < 1$, hence we only seek solutions with $\lambda \in (-\infty,0)$, which we parametrize by $\lambda = -e^{2 \pi s}$ with $s \in \mathbb{R}$, so that $\tilde{\nu} = \tanh(\pi s)$.
We find the unique solution (up to a constant $C \in \mathbb{C}$) to the boundary value problem \eqref{cond:BVP_1}--\eqref{cond:BVP_4} is
\begin{align}
S(z) &= C (z-a_1)^{-\tfrac12+is} (z-a_2)^{-\tfrac12-is} \nonumber \\
&\qquad \quad \times (z-b_1)^{-\tfrac12-is} (z-b_2)^{-\tfrac12+is}. \label{eq:BVP_soln}
\end{align}

Imposing the vanishing integral conditions to the boundary values $S^\pm(x)$ on $A$ and $B$, we find they both reduce to the condition \cref{eq:conical_zeros}.
Values of $s$ for which \cref{eq:conical_zeros} holds thus determine the spectrum of $J^\ammaG$.
Crucially, the zeros of $P_{-is-1/2}$ (as a function of $s$) form a discrete set $\{ s_n \}_{n \in \mathbb{N}}$~\cite{zhurina_tables_1966}, hence the vanishing integral conditions discretize the spectrum of $J^\ammaG$ in the range $| \tilde{\nu} | < 1$.
The values of $s_n$ do not have a simple closed form, but we can approximate them in the regimes $\eta \to 0$ (large separation) and $\eta \to 1$ (small separation).

\textbf{\textit{Logarithmic negativity.---}}For $\eta \to 0$, the zeros of the conical functions can be approximated by $s_n \approx j_{0,n}/\rho$, where $\rho := \text{arccosh}(\frac{1+\eta}{1-\eta})$.
We then determine the leading order contribution of each eigenvalue $\tilde{\nu}_n$ to the logarithmic negativity to be
\begin{align}
E_{\mathcal{N},n} \sim 2 e^{-\frac{\pi j_{0,n}}{\sqrt{\eta}}}.
\end{align}
When the lengths of the two intervals are equal, this is $E_{\mathcal{N},n} \sim 2 e^{-\pi j_{0,n} \left( 1 + \frac{d}{\ell} \right)}$.

In the small separation regime, $\eta \to 1$, the zeros of the conical functions can be approximated by $\rho s_n \approx n \pi + \text{Arg} \, B(\tfrac12, \tfrac12 is_n)$, where the second term is the complex argument of the beta function~\cite{zhurina_tables_1966,hobson_theory_1931}.
We find the leading order contribution of each eigenvalue to the logarithmic negativity is
\begin{align}
  E_{\mathcal{N},n} \sim \log_2 ( - \ln ( 1 -\eta ) ) - \log_2(n \pi^2).
\end{align}
Note that this diverges in the limit $\eta \to 1$ (or $d/\ell \to 0$).

We also calculate the leading order behavior of the total logarithmic negativity in the small separation regime by noting that the spacing between the zeros decreases as the intervals get closer, $s_{n+1} - s_n = \frac{\pi}{\rho} + \mathcal{O}(\rho^{-2})$.
Thus, we can replace the sum over $s_n$ by an integral with spectral density $\rho/\pi$, and obtain \cref{eq:lognegzerosep}.

\textbf{\textit{Negativity cores.---}}In~\cite{klco_entanglement_2023,gao_partial_2024,gao_detecting_2025}, it was shown how the eigenfunctions of $J^\ammaG$ inform the structure of entanglement between $A$ and $B$.
One can construct pairs of modes in $A$ and $B$, called negativity cores, each of which contributes $-\log_2 \tilde{\nu}_n$ to the logarithmic negativity.
The collection of cores constitutes a subsystem of $A \cup B$ containing all of the negativity.
Furthermore, the $n=1$ core is the optimal pair for accessing the negativity, in the sense that no other pair of modes in $A$ and $B$ can have larger logarithmic negativity.

The core modes are constructed from the restrictions of the eigenfunctions of $J^\ammaG$ to the intervals $A$ and $B$.
The restriction of the two eigenfunctions with eigenvalues $\pm i \tilde{\nu}_n$ to $A$ (similarly $B$) defines a two dimensional subspace of $\Gamma^T V_A^\ast$ (or $\Gamma^T V_B^\ast$) which can be identified with a mode of $A$ (or $B$), after inverting $\Gamma^T$ and normalizing the functions.
By construction, the resulting pair of modes in $A$ and $B$ has logarithmic negativity $-\log_2 \tilde{\nu}_n$.
The scalar vacuum has the special property that the restrictions to $A$ (similarly $B$) for different $n$'s yield separate, commuting modes~\cite{klco_entanglement_2023,gao_partial_2024,gao_detecting_2025}.
Hence the above construction can be applied simultaneously for all $n$ and results in the mode functions of \cref{eq:coremode_A,eq:coremode_B}.

The real and imaginary parts of these functions are the smearing functions which can be used by observers to couple to the modes' quadrature operators in \cref{eq:quadratures}.
These operators obey the canonical commutation relations, thus $(\hat{Q}_{nA},\hat{P}_{nA})$ in $A$ and $(\hat{Q}_{nB},\hat{P}_{nB})$ in $B$ for each $n$ define a pair of modes which constitutes a negativity core.

Note that for each $n$, the value of $s_n$ increases as the separation between the intervals increases.
This has the effect of decreasing the negativity between the mode pair, but it also increases the rate of oscillations of the corresponding smearing functions (seen in \cref{eq:coremode_A,eq:coremode_B}).
Hence the long-range entanglement in the field depends on high frequencies within the intervals, which explicitly demonstrates the UV-IR connection observed numerically in~\cite{klco_entanglement_2021}.

\begin{figure}[t]
  \centering
  \includegraphics[width=\linewidth]{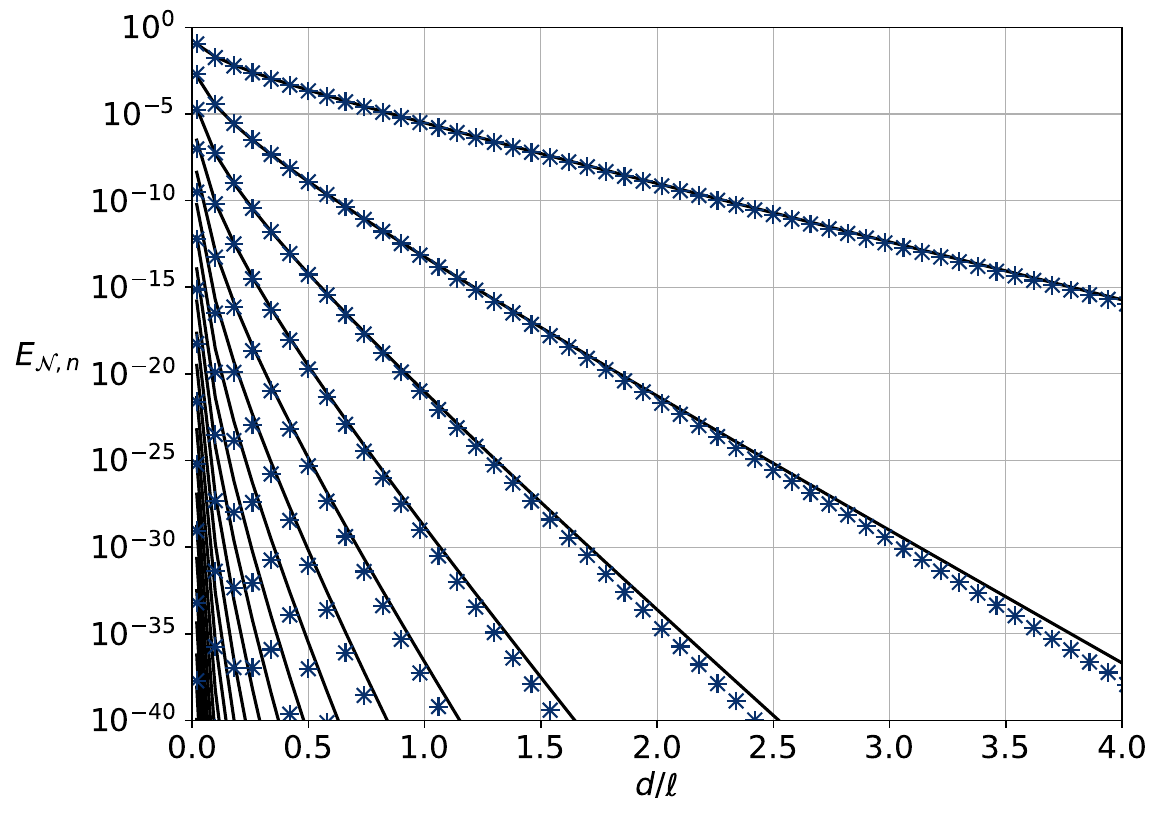}
  \caption{Numerical values of the contribution to logarithmic negativity of each eigenvalue of $J^\ammaG$. The numerical results are shown with `$\times$' or `$+$', and our analytical results with solid black lines. The parameters for this simulation were $N = 20000$, $\ell = N_A = N_B = 50$, and the separation $d$ ranged from $1$ to $4 \ell$.}
  \label{fig:numerical_negativity_permode}
\end{figure}

\textbf{\textit{Numerical validation.---}}We verify our analytical results numerically with the discrete model Hamiltonian
\begin{align}
  H = \frac{1}{2 \Delta x} \sum_{n=0}^{N-1} \left[ p_n^2 + (2 + \mu^2 \Delta x^2) q_n^2 - 2 q_n q_{n-1} \right],
\end{align}
where $q_n := \phi(x_n)$ and $p_n := \Delta x \, \pi(x_n)$ are discretized field operators with $[ q_n, p_{n'} ] = i\delta_{nn'}$.
The mass $\mu$ is introduced as a temporary regulator.

To reflect that only the derivatives $\phi'(x)$ are local observables, for an interval of sites $n \in [0,N_A]$, we apply the transformation $q' = M q$, defined by $q_0' = q_0$ and $q_n' = q_n - q_{n-1}$ for $n \geq 1$.
The new $q'$ quadratures then correspond to discrete derivatives, except for the first site $q_0'$.
To ensure that $M$ is a symplectic transformation, we define $p' = M^{-T} p$, given by $p_n' = \sum_{\ell=n}^{N_A} p_\ell$.
Finally, to obtain the subsystem $A$, we trace out the first site $(q_0',p_0')$, and are left with only discrete derivative quadratures $q'$ in the interval.
After removing $(q_0',p_0')$, we can explicitly take $\mu \to 0$.
Note that linear observables $f^T q$ transform as $f' = M^{-T} f$, hence removing $q_0'$ also removes $f_0' = \sum_{n=0}^{N_A} f_n$, which is the discrete analogue of the vanishing integral condition.

We apply these transformations (and partial trace over the first site) to two intervals of sites, $A$ and $B$, and compute the logarithmic negativity numerically for the ground state of the model.
\cref{fig:numerical_negativity_permode} shows the resulting contributions from each eigenvalue of $J^\ammaG$.
Note that there is a twofold degeneracy in each numerical eigenvalue, which is consistent with the expected equal contributions from the right- and left-moving fields.
We also see that the numerical values begin to deviate from the analytical values as the separation increases, and this occurs at a smaller separation for the modes with larger eigenvalues of $J^\ammaG$.
Both of these observations are consistent with the fact that the mode functions with smaller negativity exhibit higher frequency oscillations, which are less reliably captured by the discretized model.

\textbf{\textit{Outlook.---}}The results in this paper lend themselves to various extensions.
The framework of~\cite{hackl_bosonic_2021} has also been developed for fermionic fields, and applying our methods to that context would lead to more concrete connections with expressions for the negativity Hamiltonian in~\cite{murciano_negativity_2022}.
Further, the procedure of reformulating integral equations for calculating entanglement into boundary value problems has been developed in higher dimensions~\cite{arias_anisotropic_2017}, hence it would be natural to adapt this method for the negativity, as we have done here.
Also, in~\cite{klco_entanglement_2023,gao_partial_2024,gao_detecting_2025} it was shown that there is a rich structure of correlations between the negativity core and the remainder of the subsystem $A \cup B$ (called the ``halo''), which could be further studied analytically by finding the eigenfunctions of $J^\ammaG$ with $|\tilde{\nu}| \geq 1$.
It would also be interesting to compare with other mode decompositions~\cite{wolfNotSoNormalModeDecomposition2008}, as well as analyze the energy cost of extracting the entanglement~\cite{hackl_minimal_2019}.
We believe that the analytical approach developed here opens a path towards a deeper understanding of mixed state entanglement in quantum field theory.

\textbf{\textit{Acknowledgments.---}}JP and RHJ gratefully acknowledge support by the Wenner-Gren Foundations and the Wallenberg Initiative on Networks and Quantum Information (WINQ).
Nordita is supported in part by NordForsk.

\bibliographystyle{apsrev4-2}
\bibliography{main.bib}

\begin{thebibliography}{54}%
\makeatletter
\providecommand \@ifxundefined [1]{%
 \@ifx{#1\undefined}
}%
\providecommand \@ifnum [1]{%
 \ifnum #1\expandafter \@firstoftwo
 \else \expandafter \@secondoftwo
 \fi
}%
\providecommand \@ifx [1]{%
 \ifx #1\expandafter \@firstoftwo
 \else \expandafter \@secondoftwo
 \fi
}%
\providecommand \natexlab [1]{#1}%
\providecommand \enquote  [1]{``#1''}%
\providecommand \bibnamefont  [1]{#1}%
\providecommand \bibfnamefont [1]{#1}%
\providecommand \citenamefont [1]{#1}%
\providecommand \href@noop [0]{\@secondoftwo}%
\providecommand \href [0]{\begingroup \@sanitize@url \@href}%
\providecommand \@href[1]{\@@startlink{#1}\@@href}%
\providecommand \@@href[1]{\endgroup#1\@@endlink}%
\providecommand \@sanitize@url [0]{\catcode `\\12\catcode `\$12\catcode `\&12\catcode `\#12\catcode `\^12\catcode `\_12\catcode `\%12\relax}%
\providecommand \@@startlink[1]{}%
\providecommand \@@endlink[0]{}%
\providecommand \url  [0]{\begingroup\@sanitize@url \@url }%
\providecommand \@url [1]{\endgroup\@href {#1}{\urlprefix }}%
\providecommand \urlprefix  [0]{URL }%
\providecommand \Eprint [0]{\href }%
\providecommand \doibase [0]{https://doi.org/}%
\providecommand \selectlanguage [0]{\@gobble}%
\providecommand \bibinfo  [0]{\@secondoftwo}%
\providecommand \bibfield  [0]{\@secondoftwo}%
\providecommand \translation [1]{[#1]}%
\providecommand \BibitemOpen [0]{}%
\providecommand \bibitemStop [0]{}%
\providecommand \bibitemNoStop [0]{.\EOS\space}%
\providecommand \EOS [0]{\spacefactor3000\relax}%
\providecommand \BibitemShut  [1]{\csname bibitem#1\endcsname}%
\let\auto@bib@innerbib\@empty
\bibitem [{\citenamefont {Amico}\ \emph {et~al.}(2008)\citenamefont {Amico}, \citenamefont {Fazio}, \citenamefont {Osterloh},\ and\ \citenamefont {Vedral}}]{amico_entanglement_2008}%
  \BibitemOpen
  \bibfield  {author} {\bibinfo {author} {\bibfnamefont {L.}~\bibnamefont {Amico}}, \bibinfo {author} {\bibfnamefont {R.}~\bibnamefont {Fazio}}, \bibinfo {author} {\bibfnamefont {A.}~\bibnamefont {Osterloh}},\ and\ \bibinfo {author} {\bibfnamefont {V.}~\bibnamefont {Vedral}},\ }\href {https://doi.org/10.1103/RevModPhys.80.517} {\bibfield  {journal} {\bibinfo  {journal} {Rev. Mod. Phys.}\ }\textbf {\bibinfo {volume} {80}},\ \bibinfo {pages} {517} (\bibinfo {year} {2008})}\BibitemShut {NoStop}%
\bibitem [{\citenamefont {Eisert}\ \emph {et~al.}(2010)\citenamefont {Eisert}, \citenamefont {Cramer},\ and\ \citenamefont {Plenio}}]{eisert_area_2010}%
  \BibitemOpen
  \bibfield  {author} {\bibinfo {author} {\bibfnamefont {J.}~\bibnamefont {Eisert}}, \bibinfo {author} {\bibfnamefont {M.}~\bibnamefont {Cramer}},\ and\ \bibinfo {author} {\bibfnamefont {M.~B.}\ \bibnamefont {Plenio}},\ }\href {https://doi.org/10.1103/RevModPhys.82.277} {\bibfield  {journal} {\bibinfo  {journal} {Rev. Mod. Phys.}\ }\textbf {\bibinfo {volume} {82}},\ \bibinfo {pages} {277} (\bibinfo {year} {2010})}\BibitemShut {NoStop}%
\bibitem [{\citenamefont {Ryu}\ and\ \citenamefont {Takayanagi}(2006)}]{ryu_holographic_2006}%
  \BibitemOpen
  \bibfield  {author} {\bibinfo {author} {\bibfnamefont {S.}~\bibnamefont {Ryu}}\ and\ \bibinfo {author} {\bibfnamefont {T.}~\bibnamefont {Takayanagi}},\ }\href {https://doi.org/10.1103/PhysRevLett.96.181602} {\bibfield  {journal} {\bibinfo  {journal} {Phys. Rev. Lett.}\ }\textbf {\bibinfo {volume} {96}},\ \bibinfo {pages} {181602} (\bibinfo {year} {2006})}\BibitemShut {NoStop}%
\bibitem [{\citenamefont {Swingle}(2012)}]{swingle_entanglement_2012}%
  \BibitemOpen
  \bibfield  {author} {\bibinfo {author} {\bibfnamefont {B.}~\bibnamefont {Swingle}},\ }\href {https://doi.org/10.1103/PhysRevD.86.065007} {\bibfield  {journal} {\bibinfo  {journal} {Phys. Rev. D}\ }\textbf {\bibinfo {volume} {86}},\ \bibinfo {pages} {065007} (\bibinfo {year} {2012})}\BibitemShut {NoStop}%
\bibitem [{\citenamefont {Almheiri}\ \emph {et~al.}(2019)\citenamefont {Almheiri}, \citenamefont {Engelhardt}, \citenamefont {Marolf},\ and\ \citenamefont {Maxfield}}]{almheiri_entropy_2019}%
  \BibitemOpen
  \bibfield  {author} {\bibinfo {author} {\bibfnamefont {A.}~\bibnamefont {Almheiri}}, \bibinfo {author} {\bibfnamefont {N.}~\bibnamefont {Engelhardt}}, \bibinfo {author} {\bibfnamefont {D.}~\bibnamefont {Marolf}},\ and\ \bibinfo {author} {\bibfnamefont {H.}~\bibnamefont {Maxfield}},\ }\href {https://doi.org/10.1007/JHEP12(2019)063} {\bibfield  {journal} {\bibinfo  {journal} {JHEP}\ }\textbf {\bibinfo {volume} {2019}}\bibinfo  {number} { (12)},\ \bibinfo {pages} {63}}\BibitemShut {NoStop}%
\bibitem [{\citenamefont {Penington}(2020)}]{penington_entanglement_2020}%
  \BibitemOpen
\bibfield  {number} {  }\bibfield  {author} {\bibinfo {author} {\bibfnamefont {G.}~\bibnamefont {Penington}},\ }\href {https://doi.org/10.1007/JHEP09(2020)002} {\bibfield  {journal} {\bibinfo  {journal} {JHEP}\ }\textbf {\bibinfo {volume} {2020}}\bibinfo  {number} { (9)},\ \bibinfo {pages} {2}}\BibitemShut {NoStop}%
\bibitem [{\citenamefont {Van~Raamsdonk}(2010)}]{vanraamsdonk_building_2010}%
  \BibitemOpen
\bibfield  {number} {  }\bibfield  {author} {\bibinfo {author} {\bibfnamefont {M.}~\bibnamefont {Van~Raamsdonk}},\ }\href {https://doi.org/10.1142/S0218271810018529} {\bibfield  {journal} {\bibinfo  {journal} {Gen. Rel. and Grav.}\ }\textbf {\bibinfo {volume} {42}},\ \bibinfo {pages} {2323} (\bibinfo {year} {2010})}\BibitemShut {NoStop}%
\bibitem [{\citenamefont {Jacobson}(2016)}]{jacobson_entanglement_2016}%
  \BibitemOpen
  \bibfield  {author} {\bibinfo {author} {\bibfnamefont {T.}~\bibnamefont {Jacobson}},\ }\href {https://doi.org/10.1103/PhysRevLett.116.201101} {\bibfield  {journal} {\bibinfo  {journal} {Phys. Rev. Lett.}\ }\textbf {\bibinfo {volume} {116}},\ \bibinfo {pages} {201101} (\bibinfo {year} {2016})}\BibitemShut {NoStop}%
\bibitem [{\citenamefont {Cao}\ \emph {et~al.}(2017)\citenamefont {Cao}, \citenamefont {Carroll},\ and\ \citenamefont {Michalakis}}]{cao_space_2017}%
  \BibitemOpen
  \bibfield  {author} {\bibinfo {author} {\bibfnamefont {C.}~\bibnamefont {Cao}}, \bibinfo {author} {\bibfnamefont {S.~M.}\ \bibnamefont {Carroll}},\ and\ \bibinfo {author} {\bibfnamefont {S.}~\bibnamefont {Michalakis}},\ }\href {https://doi.org/10.1103/PhysRevD.95.024031} {\bibfield  {journal} {\bibinfo  {journal} {Phys. Rev. D}\ }\textbf {\bibinfo {volume} {95}},\ \bibinfo {pages} {024031} (\bibinfo {year} {2017})}\BibitemShut {NoStop}%
\bibitem [{\citenamefont {Sorkin}(1983)}]{sorkin_entropy_1983}%
  \BibitemOpen
  \bibfield  {author} {\bibinfo {author} {\bibfnamefont {R.~D.}\ \bibnamefont {Sorkin}},\ }in\ \href@noop {} {\emph {\bibinfo {booktitle} {Tenth International Conference on General Relativity and Gravitation (held Padova, 4-9 July, 1983), Contributed Papers}}},\ Vol.~\bibinfo {volume} {2}\ (\bibinfo {year} {1983})\ pp.\ \bibinfo {pages} {734--736},\ \Eprint {https://arxiv.org/abs/1402.3589} {arXiv:1402.3589 [gr-qc]} \BibitemShut {NoStop}%
\bibitem [{\citenamefont {Bombelli}\ \emph {et~al.}(1986)\citenamefont {Bombelli}, \citenamefont {Koul}, \citenamefont {Lee},\ and\ \citenamefont {Sorkin}}]{bombelli_quantum_1986}%
  \BibitemOpen
  \bibfield  {author} {\bibinfo {author} {\bibfnamefont {L.}~\bibnamefont {Bombelli}}, \bibinfo {author} {\bibfnamefont {R.~K.}\ \bibnamefont {Koul}}, \bibinfo {author} {\bibfnamefont {J.}~\bibnamefont {Lee}},\ and\ \bibinfo {author} {\bibfnamefont {R.~D.}\ \bibnamefont {Sorkin}},\ }\href {https://doi.org/10.1103/PhysRevD.34.373} {\bibfield  {journal} {\bibinfo  {journal} {Phys. Rev. D}\ }\textbf {\bibinfo {volume} {34}},\ \bibinfo {pages} {373} (\bibinfo {year} {1986})}\BibitemShut {NoStop}%
\bibitem [{\citenamefont {Srednicki}(1993)}]{srednicki_entropy_1993}%
  \BibitemOpen
  \bibfield  {author} {\bibinfo {author} {\bibfnamefont {M.}~\bibnamefont {Srednicki}},\ }\href {https://doi.org/10.1103/PhysRevLett.71.666} {\bibfield  {journal} {\bibinfo  {journal} {Phys. Rev. Lett.}\ }\textbf {\bibinfo {volume} {71}},\ \bibinfo {pages} {666} (\bibinfo {year} {1993})}\BibitemShut {NoStop}%
\bibitem [{\citenamefont {Callan}\ and\ \citenamefont {Wilczek}(1994)}]{callan_geometric_1994}%
  \BibitemOpen
  \bibfield  {author} {\bibinfo {author} {\bibfnamefont {C.}~\bibnamefont {Callan}}\ and\ \bibinfo {author} {\bibfnamefont {F.}~\bibnamefont {Wilczek}},\ }\href {https://doi.org/https://doi.org/10.1016/0370-2693(94)91007-3} {\bibfield  {journal} {\bibinfo  {journal} {Phys. Lett. B}\ }\textbf {\bibinfo {volume} {333}},\ \bibinfo {pages} {55} (\bibinfo {year} {1994})}\BibitemShut {NoStop}%
\bibitem [{\citenamefont {Solodukhin}(2011)}]{solodukhin_entanglement_2011}%
  \BibitemOpen
  \bibfield  {author} {\bibinfo {author} {\bibfnamefont {S.~N.}\ \bibnamefont {Solodukhin}},\ }\href {https://doi.org/10.12942/lrr-2011-8} {\bibfield  {journal} {\bibinfo  {journal} {Liv. Rev. Rel.}\ }\textbf {\bibinfo {volume} {14}},\ \bibinfo {pages} {8} (\bibinfo {year} {2011})}\BibitemShut {NoStop}%
\bibitem [{\citenamefont {Redhead}(1995)}]{redhead_ado_1995}%
  \BibitemOpen
  \bibfield  {author} {\bibinfo {author} {\bibfnamefont {M.}~\bibnamefont {Redhead}},\ }\href {https://doi.org/10.1007/BF02054660} {\bibfield  {journal} {\bibinfo  {journal} {Found. Phys.}\ }\textbf {\bibinfo {volume} {25}},\ \bibinfo {pages} {123} (\bibinfo {year} {1995})}\BibitemShut {NoStop}%
\bibitem [{\citenamefont {Witten}(2018)}]{witten_aps_2018}%
  \BibitemOpen
  \bibfield  {author} {\bibinfo {author} {\bibfnamefont {E.}~\bibnamefont {Witten}},\ }\href {https://doi.org/10.1103/RevModPhys.90.045003} {\bibfield  {journal} {\bibinfo  {journal} {Rev. Mod. Phys.}\ }\textbf {\bibinfo {volume} {90}},\ \bibinfo {pages} {045003} (\bibinfo {year} {2018})}\BibitemShut {NoStop}%
\bibitem [{\citenamefont {Halvorson}\ and\ \citenamefont {Clifton}(2002)}]{halvorson_particles_2002}%
  \BibitemOpen
  \bibfield  {author} {\bibinfo {author} {\bibfnamefont {H.}~\bibnamefont {Halvorson}}\ and\ \bibinfo {author} {\bibfnamefont {R.}~\bibnamefont {Clifton}},\ }\href {https://doi.org/10.1086/338939} {\bibfield  {journal} {\bibinfo  {journal} {Phil. Sci.}\ }\textbf {\bibinfo {volume} {69}},\ \bibinfo {pages} {1} (\bibinfo {year} {2002})}\BibitemShut {NoStop}%
\bibitem [{\citenamefont {Papageorgiou}\ and\ \citenamefont {Pye}(2019)}]{papageorgiou_impact_2019}%
  \BibitemOpen
  \bibfield  {author} {\bibinfo {author} {\bibfnamefont {M.}~\bibnamefont {Papageorgiou}}\ and\ \bibinfo {author} {\bibfnamefont {J.}~\bibnamefont {Pye}},\ }\href {https://doi.org/10.1088/1751-8121/ab3593} {\bibfield  {journal} {\bibinfo  {journal} {J. Phys. A: Math. Th.}\ }\textbf {\bibinfo {volume} {52}},\ \bibinfo {pages} {375304} (\bibinfo {year} {2019})}\BibitemShut {NoStop}%
\bibitem [{\citenamefont {Summers}\ and\ \citenamefont {Werner}(1987)}]{summers_maximal_1987}%
  \BibitemOpen
  \bibfield  {author} {\bibinfo {author} {\bibfnamefont {S.~J.}\ \bibnamefont {Summers}}\ and\ \bibinfo {author} {\bibfnamefont {R.}~\bibnamefont {Werner}},\ }\href {https://doi.org/10.1007/BF01207366} {\bibfield  {journal} {\bibinfo  {journal} {Comm. Math. Phys.}\ }\textbf {\bibinfo {volume} {110}},\ \bibinfo {pages} {247} (\bibinfo {year} {1987})}\BibitemShut {NoStop}%
\bibitem [{\citenamefont {Verch}\ and\ \citenamefont {Werner}(2005)}]{verch_distillability_2005}%
  \BibitemOpen
  \bibfield  {author} {\bibinfo {author} {\bibfnamefont {R.}~\bibnamefont {Verch}}\ and\ \bibinfo {author} {\bibfnamefont {R.~F.}\ \bibnamefont {Werner}},\ }\href {https://doi.org/10.1142/S0129055X05002364} {\bibfield  {journal} {\bibinfo  {journal} {Rev. Math. Phys.}\ }\textbf {\bibinfo {volume} {17}},\ \bibinfo {pages} {545} (\bibinfo {year} {2005})}\BibitemShut {NoStop}%
\bibitem [{\citenamefont {Valentini}(1991)}]{valentiniNonlocalCorrelationsQuantum1991}%
  \BibitemOpen
  \bibfield  {author} {\bibinfo {author} {\bibfnamefont {A.}~\bibnamefont {Valentini}},\ }\href {https://doi.org/10.1016/0375-9601(91)90952-5} {\bibfield  {journal} {\bibinfo  {journal} {Phys. Lett. A}\ }\textbf {\bibinfo {volume} {153}},\ \bibinfo {pages} {321} (\bibinfo {year} {1991})}\BibitemShut {NoStop}%
\bibitem [{\citenamefont {Reznik}(2003)}]{reznikEntanglementVacuum2003}%
  \BibitemOpen
  \bibfield  {author} {\bibinfo {author} {\bibfnamefont {B.}~\bibnamefont {Reznik}},\ }\href {https://doi.org/10.1023/A:1022875910744} {\bibfield  {journal} {\bibinfo  {journal} {Found. Phys.}\ }\textbf {\bibinfo {volume} {33}},\ \bibinfo {pages} {167} (\bibinfo {year} {2003})}\BibitemShut {NoStop}%
\bibitem [{\citenamefont {Steeg}\ and\ \citenamefont {Menicucci}(2009)}]{steegEntanglingPowerExpanding2009}%
  \BibitemOpen
  \bibfield  {author} {\bibinfo {author} {\bibfnamefont {G.~V.}\ \bibnamefont {Steeg}}\ and\ \bibinfo {author} {\bibfnamefont {N.~C.}\ \bibnamefont {Menicucci}},\ }\href {https://doi.org/10.1103/PhysRevD.79.044027} {\bibfield  {journal} {\bibinfo  {journal} {Phys. Rev. D}\ }\textbf {\bibinfo {volume} {79}},\ \bibinfo {pages} {044027} (\bibinfo {year} {2009})}\BibitemShut {NoStop}%
\bibitem [{\citenamefont {{Mart{\'i}n-Mart{\'i}nez}}\ and\ \citenamefont {Menicucci}(2014)}]{martin-martinezEntanglementCurvedSpacetimes2014}%
  \BibitemOpen
  \bibfield  {author} {\bibinfo {author} {\bibfnamefont {E.}~\bibnamefont {{Mart{\'i}n-Mart{\'i}nez}}}\ and\ \bibinfo {author} {\bibfnamefont {N.~C.}\ \bibnamefont {Menicucci}},\ }\href {https://doi.org/10.1088/0264-9381/31/21/214001} {\bibfield  {journal} {\bibinfo  {journal} {Class. Quant. Grav.}\ }\textbf {\bibinfo {volume} {31}},\ \bibinfo {pages} {214001} (\bibinfo {year} {2014})}\BibitemShut {NoStop}%
\bibitem [{\citenamefont {Carib{\'e}}\ \emph {et~al.}(2023)\citenamefont {Carib{\'e}}, \citenamefont {Jonsson}, \citenamefont {Casals}, \citenamefont {Kempf},\ and\ \citenamefont {{Mart{\'i}n-Mart{\'i}nez}}}]{caribeLensingVacuumEntanglement2023}%
  \BibitemOpen
  \bibfield  {author} {\bibinfo {author} {\bibfnamefont {J.~G.~A.}\ \bibnamefont {Carib{\'e}}}, \bibinfo {author} {\bibfnamefont {R.~H.}\ \bibnamefont {Jonsson}}, \bibinfo {author} {\bibfnamefont {M.}~\bibnamefont {Casals}}, \bibinfo {author} {\bibfnamefont {A.}~\bibnamefont {Kempf}},\ and\ \bibinfo {author} {\bibfnamefont {E.}~\bibnamefont {{Mart{\'i}n-Mart{\'i}nez}}},\ }\href {https://doi.org/10.1103/PhysRevD.108.025016} {\bibfield  {journal} {\bibinfo  {journal} {Phys. Rev. D}\ }\textbf {\bibinfo {volume} {108}},\ \bibinfo {pages} {025016} (\bibinfo {year} {2023})}\BibitemShut {NoStop}%
\bibitem [{\citenamefont {Calabrese}\ and\ \citenamefont {Cardy}(2004)}]{calabrese_entanglement_2004}%
  \BibitemOpen
  \bibfield  {author} {\bibinfo {author} {\bibfnamefont {P.}~\bibnamefont {Calabrese}}\ and\ \bibinfo {author} {\bibfnamefont {J.}~\bibnamefont {Cardy}},\ }\href {https://doi.org/10.1088/1742-5468/2004/06/P06002} {\bibfield  {journal} {\bibinfo  {journal} {J. Stat. Mech.: Th. Exp.}\ }\textbf {\bibinfo {volume} {2004}},\ \bibinfo {pages} {P06002} (\bibinfo {year} {2004})}\BibitemShut {NoStop}%
\bibitem [{\citenamefont {Calabrese}\ and\ \citenamefont {Cardy}(2009)}]{calabrese_entanglement_2009}%
  \BibitemOpen
  \bibfield  {author} {\bibinfo {author} {\bibfnamefont {P.}~\bibnamefont {Calabrese}}\ and\ \bibinfo {author} {\bibfnamefont {J.}~\bibnamefont {Cardy}},\ }\href {https://doi.org/10.1088/1751-8113/42/50/504005} {\bibfield  {journal} {\bibinfo  {journal} {J. Phys. A: Math. Th.}\ }\textbf {\bibinfo {volume} {42}},\ \bibinfo {pages} {504005} (\bibinfo {year} {2009})}\BibitemShut {NoStop}%
\bibitem [{\citenamefont {Casini}\ and\ \citenamefont {Huerta}(2009)}]{casini_entanglement_2009}%
  \BibitemOpen
  \bibfield  {author} {\bibinfo {author} {\bibfnamefont {H.}~\bibnamefont {Casini}}\ and\ \bibinfo {author} {\bibfnamefont {M.}~\bibnamefont {Huerta}},\ }\href {https://doi.org/10.1088/1751-8113/42/50/504007} {\bibfield  {journal} {\bibinfo  {journal} {J. Phys. A: Math. Th.}\ }\textbf {\bibinfo {volume} {42}},\ \bibinfo {pages} {504007} (\bibinfo {year} {2009})}\BibitemShut {NoStop}%
\bibitem [{\citenamefont {Casini}\ and\ \citenamefont {Huerta}(2022)}]{casini_lectures_2022}%
  \BibitemOpen
  \bibfield  {author} {\bibinfo {author} {\bibfnamefont {H.}~\bibnamefont {Casini}}\ and\ \bibinfo {author} {\bibfnamefont {M.}~\bibnamefont {Huerta}},\ }\href@noop {} {\bibfield  {journal} {\bibinfo  {journal} {arXiv:2201.13310}\ } (\bibinfo {year} {2022})},\ \Eprint {https://arxiv.org/abs/2201.13310} {arXiv:2201.13310} \BibitemShut {NoStop}%
\bibitem [{\citenamefont {Bianchi}\ and\ \citenamefont {Satz}(2019)}]{bianchiEntropySubalgebraObservables2019}%
  \BibitemOpen
  \bibfield  {author} {\bibinfo {author} {\bibfnamefont {E.}~\bibnamefont {Bianchi}}\ and\ \bibinfo {author} {\bibfnamefont {A.}~\bibnamefont {Satz}},\ }\href {https://doi.org/10.1103/PhysRevD.99.085001} {\bibfield  {journal} {\bibinfo  {journal} {Phys. Rev. D}\ }\textbf {\bibinfo {volume} {99}},\ \bibinfo {pages} {085001} (\bibinfo {year} {2019})}\BibitemShut {NoStop}%
\bibitem [{\citenamefont {Horodecki}\ \emph {et~al.}(2009)\citenamefont {Horodecki}, \citenamefont {Horodecki}, \citenamefont {Horodecki},\ and\ \citenamefont {Horodecki}}]{horodecki_quantum_2009}%
  \BibitemOpen
  \bibfield  {author} {\bibinfo {author} {\bibfnamefont {R.}~\bibnamefont {Horodecki}}, \bibinfo {author} {\bibfnamefont {P.}~\bibnamefont {Horodecki}}, \bibinfo {author} {\bibfnamefont {M.}~\bibnamefont {Horodecki}},\ and\ \bibinfo {author} {\bibfnamefont {K.}~\bibnamefont {Horodecki}},\ }\href {https://doi.org/10.1103/RevModPhys.81.865} {\bibfield  {journal} {\bibinfo  {journal} {Rev. Mod. Phys.}\ }\textbf {\bibinfo {volume} {81}},\ \bibinfo {pages} {865} (\bibinfo {year} {2009})}\BibitemShut {NoStop}%
\bibitem [{\citenamefont {Serafini}(2023)}]{serafini_quantum_2023}%
  \BibitemOpen
  \bibfield  {author} {\bibinfo {author} {\bibfnamefont {A.}~\bibnamefont {Serafini}},\ }\href@noop {} {\emph {\bibinfo {title} {Quantum continuous variables: a primer of theoretical methods}}},\ \bibinfo {edition} {2nd}\ ed.\ (\bibinfo  {publisher} {CRC press},\ \bibinfo {year} {2023})\BibitemShut {NoStop}%
\bibitem [{\citenamefont {Marcovitch}\ \emph {et~al.}(2009)\citenamefont {Marcovitch}, \citenamefont {Retzker}, \citenamefont {Plenio},\ and\ \citenamefont {Reznik}}]{marcovitch_critical_2009}%
  \BibitemOpen
  \bibfield  {author} {\bibinfo {author} {\bibfnamefont {S.}~\bibnamefont {Marcovitch}}, \bibinfo {author} {\bibfnamefont {A.}~\bibnamefont {Retzker}}, \bibinfo {author} {\bibfnamefont {M.~B.}\ \bibnamefont {Plenio}},\ and\ \bibinfo {author} {\bibfnamefont {B.}~\bibnamefont {Reznik}},\ }\href {https://doi.org/10.1103/PhysRevA.80.012325} {\bibfield  {journal} {\bibinfo  {journal} {Phys. Rev. A}\ }\textbf {\bibinfo {volume} {80}},\ \bibinfo {pages} {012325} (\bibinfo {year} {2009})}\BibitemShut {NoStop}%
\bibitem [{\citenamefont {Klco}\ and\ \citenamefont {Savage}(2021)}]{klco_entanglement_2021}%
  \BibitemOpen
  \bibfield  {author} {\bibinfo {author} {\bibfnamefont {N.}~\bibnamefont {Klco}}\ and\ \bibinfo {author} {\bibfnamefont {M.~J.}\ \bibnamefont {Savage}},\ }\href {https://doi.org/10.1103/PhysRevLett.127.211602} {\bibfield  {journal} {\bibinfo  {journal} {Phys. Rev. Lett.}\ }\textbf {\bibinfo {volume} {127}},\ \bibinfo {pages} {211602} (\bibinfo {year} {2021})}\BibitemShut {NoStop}%
\bibitem [{\citenamefont {Calabrese}\ \emph {et~al.}(2012)\citenamefont {Calabrese}, \citenamefont {Cardy},\ and\ \citenamefont {Tonni}}]{calabrese_entanglement_2012}%
  \BibitemOpen
  \bibfield  {author} {\bibinfo {author} {\bibfnamefont {P.}~\bibnamefont {Calabrese}}, \bibinfo {author} {\bibfnamefont {J.}~\bibnamefont {Cardy}},\ and\ \bibinfo {author} {\bibfnamefont {E.}~\bibnamefont {Tonni}},\ }\href {https://doi.org/10.1103/PhysRevLett.109.130502} {\bibfield  {journal} {\bibinfo  {journal} {Phys. Rev. Lett.}\ }\textbf {\bibinfo {volume} {109}},\ \bibinfo {pages} {130502} (\bibinfo {year} {2012})}\BibitemShut {NoStop}%
\bibitem [{\citenamefont {Calabrese}\ \emph {et~al.}(2013)\citenamefont {Calabrese}, \citenamefont {Cardy},\ and\ \citenamefont {Tonni}}]{calabrese_entanglement_2013}%
  \BibitemOpen
  \bibfield  {author} {\bibinfo {author} {\bibfnamefont {P.}~\bibnamefont {Calabrese}}, \bibinfo {author} {\bibfnamefont {J.}~\bibnamefont {Cardy}},\ and\ \bibinfo {author} {\bibfnamefont {E.}~\bibnamefont {Tonni}},\ }\href {https://doi.org/10.1088/1742-5468/2013/02/P02008} {\bibfield  {journal} {\bibinfo  {journal} {J. Stat. Mech.: Th. Exp.}\ }\textbf {\bibinfo {volume} {2013}},\ \bibinfo {pages} {P02008} (\bibinfo {year} {2013})}\BibitemShut {NoStop}%
\bibitem [{\citenamefont {Arias}\ \emph {et~al.}(2026)\citenamefont {Arias}, \citenamefont {Huerta}, \citenamefont {Rotaru},\ and\ \citenamefont {Tonni}}]{arias_entanglement_2026}%
  \BibitemOpen
  \bibfield  {author} {\bibinfo {author} {\bibfnamefont {M.}~\bibnamefont {Arias}}, \bibinfo {author} {\bibfnamefont {M.}~\bibnamefont {Huerta}}, \bibinfo {author} {\bibfnamefont {A.}~\bibnamefont {Rotaru}},\ and\ \bibinfo {author} {\bibfnamefont {E.}~\bibnamefont {Tonni}},\ }\href@noop {} {\bibfield  {journal} {\bibinfo  {journal} {arXiv:2601.04995}\ } (\bibinfo {year} {2026})},\ \Eprint {https://arxiv.org/abs/2601.04995} {arXiv:2601.04995} \BibitemShut {NoStop}%
\bibitem [{\citenamefont {Zych}\ \emph {et~al.}(2010)\citenamefont {Zych}, \citenamefont {Costa}, \citenamefont {Kofler},\ and\ \citenamefont {Brukner}}]{zych_entanglement_2010}%
  \BibitemOpen
  \bibfield  {author} {\bibinfo {author} {\bibfnamefont {M.}~\bibnamefont {Zych}}, \bibinfo {author} {\bibfnamefont {F.}~\bibnamefont {Costa}}, \bibinfo {author} {\bibfnamefont {J.}~\bibnamefont {Kofler}},\ and\ \bibinfo {author} {\bibfnamefont {C.}~\bibnamefont {Brukner}},\ }\href {https://doi.org/10.1103/PhysRevD.81.125019} {\bibfield  {journal} {\bibinfo  {journal} {Phys. Rev. D}\ }\textbf {\bibinfo {volume} {81}},\ \bibinfo {pages} {125019} (\bibinfo {year} {2010})}\BibitemShut {NoStop}%
\bibitem [{\citenamefont {Klco}\ \emph {et~al.}(2023)\citenamefont {Klco}, \citenamefont {Beck},\ and\ \citenamefont {Savage}}]{klco_entanglement_2023}%
  \BibitemOpen
  \bibfield  {author} {\bibinfo {author} {\bibfnamefont {N.}~\bibnamefont {Klco}}, \bibinfo {author} {\bibfnamefont {D.~H.}\ \bibnamefont {Beck}},\ and\ \bibinfo {author} {\bibfnamefont {M.~J.}\ \bibnamefont {Savage}},\ }\href {https://doi.org/10.1103/PhysRevA.107.012415} {\bibfield  {journal} {\bibinfo  {journal} {Phys. Rev. A}\ }\textbf {\bibinfo {volume} {107}},\ \bibinfo {pages} {012415} (\bibinfo {year} {2023})}\BibitemShut {NoStop}%
\bibitem [{\citenamefont {Gao}\ and\ \citenamefont {Klco}(2024)}]{gao_partial_2024}%
  \BibitemOpen
  \bibfield  {author} {\bibinfo {author} {\bibfnamefont {B.}~\bibnamefont {Gao}}\ and\ \bibinfo {author} {\bibfnamefont {N.}~\bibnamefont {Klco}},\ }\href {https://doi.org/10.1103/PhysRevA.109.062413} {\bibfield  {journal} {\bibinfo  {journal} {Phys. Rev. A}\ }\textbf {\bibinfo {volume} {109}},\ \bibinfo {pages} {062413} (\bibinfo {year} {2024})}\BibitemShut {NoStop}%
\bibitem [{\citenamefont {Gao}\ and\ \citenamefont {Klco}(2025)}]{gao_detecting_2025}%
  \BibitemOpen
  \bibfield  {author} {\bibinfo {author} {\bibfnamefont {B.}~\bibnamefont {Gao}}\ and\ \bibinfo {author} {\bibfnamefont {N.}~\bibnamefont {Klco}},\ }\href {https://doi.org/10.1103/m9w1-ppqz} {\bibfield  {journal} {\bibinfo  {journal} {Phys. Rev. A}\ }\textbf {\bibinfo {volume} {112}},\ \bibinfo {pages} {012430} (\bibinfo {year} {2025})}\BibitemShut {NoStop}%
\bibitem [{\citenamefont {Agullo}\ \emph {et~al.}(2025{\natexlab{a}})\citenamefont {Agullo}, \citenamefont {Bonga}, \citenamefont {Mart{\'\i}n-Mart{\'\i}nez}, \citenamefont {Nadal-Gisbert}, \citenamefont {Perche}, \citenamefont {Polo-G{\'o}mez}, \citenamefont {Ribes-Metidieri},\ and\ \citenamefont {Torres}}]{agullo_multimode_2025}%
  \BibitemOpen
  \bibfield  {author} {\bibinfo {author} {\bibfnamefont {I.}~\bibnamefont {Agullo}}, \bibinfo {author} {\bibfnamefont {B.}~\bibnamefont {Bonga}}, \bibinfo {author} {\bibfnamefont {E.}~\bibnamefont {Mart{\'\i}n-Mart{\'\i}nez}}, \bibinfo {author} {\bibfnamefont {S.}~\bibnamefont {Nadal-Gisbert}}, \bibinfo {author} {\bibfnamefont {T.~R.}\ \bibnamefont {Perche}}, \bibinfo {author} {\bibfnamefont {J.}~\bibnamefont {Polo-G{\'o}mez}}, \bibinfo {author} {\bibfnamefont {P.}~\bibnamefont {Ribes-Metidieri}},\ and\ \bibinfo {author} {\bibfnamefont {B.~d.~S.}\ \bibnamefont {Torres}},\ }\href {https://doi.org/10.1103/PhysRevD.111.085013} {\bibfield  {journal} {\bibinfo  {journal} {Phys. Rev. D}\ }\textbf {\bibinfo {volume} {111}},\ \bibinfo {pages} {085013} (\bibinfo {year} {2025}{\natexlab{a}})}\BibitemShut {NoStop}%
\bibitem [{\citenamefont {Botero}\ and\ \citenamefont {Reznik}(2003)}]{boteroModeWiseEntanglementGaussian2003}%
  \BibitemOpen
  \bibfield  {author} {\bibinfo {author} {\bibfnamefont {A.}~\bibnamefont {Botero}}\ and\ \bibinfo {author} {\bibfnamefont {B.}~\bibnamefont {Reznik}},\ }\bibfield  {journal} {\bibinfo  {journal} {Phys. Rev. A}\ }\textbf {\bibinfo {volume} {67}},\ \href {https://doi.org/10.1103/PhysRevA.67.052311} {10.1103/PhysRevA.67.052311} (\bibinfo {year} {2003})\BibitemShut {NoStop}%
\bibitem [{\citenamefont {Botero}\ and\ \citenamefont {Reznik}(2004)}]{boteroSpatialStructuresLocalization2004a}%
  \BibitemOpen
  \bibfield  {author} {\bibinfo {author} {\bibfnamefont {A.}~\bibnamefont {Botero}}\ and\ \bibinfo {author} {\bibfnamefont {B.}~\bibnamefont {Reznik}},\ }\href {https://doi.org/10.1103/PhysRevA.70.052329} {\bibfield  {journal} {\bibinfo  {journal} {Phys. Rev. A}\ }\textbf {\bibinfo {volume} {70}},\ \bibinfo {pages} {052329} (\bibinfo {year} {2004})}\BibitemShut {NoStop}%
\bibitem [{\citenamefont {Hackl}\ and\ \citenamefont {Jonsson}(2019)}]{hackl_minimal_2019}%
  \BibitemOpen
  \bibfield  {author} {\bibinfo {author} {\bibfnamefont {L.}~\bibnamefont {Hackl}}\ and\ \bibinfo {author} {\bibfnamefont {R.~H.}\ \bibnamefont {Jonsson}},\ }\href {https://doi.org/10.22331/q-2019-07-15-165} {\bibfield  {journal} {\bibinfo  {journal} {Quantum}\ }\textbf {\bibinfo {volume} {3}},\ \bibinfo {pages} {165} (\bibinfo {year} {2019})}\BibitemShut {NoStop}%
\bibitem [{\citenamefont {Agullo}\ \emph {et~al.}(2025{\natexlab{b}})\citenamefont {Agullo}, \citenamefont {Mart{\'\i}n-Mart{\'\i}nez}, \citenamefont {Nadal-Gisbert}, \citenamefont {Ribes-Metidieri},\ and\ \citenamefont {Yamaguchi}}]{agulloCorrelationEntanglementPartners2025}%
  \BibitemOpen
  \bibfield  {author} {\bibinfo {author} {\bibfnamefont {I.}~\bibnamefont {Agullo}}, \bibinfo {author} {\bibfnamefont {E.}~\bibnamefont {Mart{\'\i}n-Mart{\'\i}nez}}, \bibinfo {author} {\bibfnamefont {S.}~\bibnamefont {Nadal-Gisbert}}, \bibinfo {author} {\bibfnamefont {P.}~\bibnamefont {Ribes-Metidieri}},\ and\ \bibinfo {author} {\bibfnamefont {K.}~\bibnamefont {Yamaguchi}},\ }\href@noop {} {\bibfield  {journal} {\bibinfo  {journal} {arXiv:2512.11055}\ } (\bibinfo {year} {2025}{\natexlab{b}})},\ \Eprint {https://arxiv.org/abs/2512.11055} {arXiv:2512.11055} \BibitemShut {NoStop}%
\bibitem [{\citenamefont {Hackl}\ and\ \citenamefont {Bianchi}(2021)}]{hackl_bosonic_2021}%
  \BibitemOpen
  \bibfield  {author} {\bibinfo {author} {\bibfnamefont {L.}~\bibnamefont {Hackl}}\ and\ \bibinfo {author} {\bibfnamefont {E.}~\bibnamefont {Bianchi}},\ }\href {https://doi.org/10.21468/SciPostPhysCore.4.3.025} {\bibfield  {journal} {\bibinfo  {journal} {SciPost Physics Core}\ }\textbf {\bibinfo {volume} {4}},\ \bibinfo {pages} {025} (\bibinfo {year} {2021})}\BibitemShut {NoStop}%
\bibitem [{\citenamefont {Murciano}\ \emph {et~al.}(2022)\citenamefont {Murciano}, \citenamefont {Vitale}, \citenamefont {Dalmonte},\ and\ \citenamefont {Calabrese}}]{murciano_negativity_2022}%
  \BibitemOpen
  \bibfield  {author} {\bibinfo {author} {\bibfnamefont {S.}~\bibnamefont {Murciano}}, \bibinfo {author} {\bibfnamefont {V.}~\bibnamefont {Vitale}}, \bibinfo {author} {\bibfnamefont {M.}~\bibnamefont {Dalmonte}},\ and\ \bibinfo {author} {\bibfnamefont {P.}~\bibnamefont {Calabrese}},\ }\href {https://doi.org/10.1103/PhysRevLett.128.140502} {\bibfield  {journal} {\bibinfo  {journal} {Phys. Rev. Lett.}\ }\textbf {\bibinfo {volume} {128}},\ \bibinfo {pages} {140502} (\bibinfo {year} {2022})}\BibitemShut {NoStop}%
\bibitem [{\citenamefont {Arias}\ \emph {et~al.}(2018)\citenamefont {Arias}, \citenamefont {Casini}, \citenamefont {Huerta},\ and\ \citenamefont {Pontello}}]{arias_entropy_2018}%
  \BibitemOpen
  \bibfield  {author} {\bibinfo {author} {\bibfnamefont {R.~E.}\ \bibnamefont {Arias}}, \bibinfo {author} {\bibfnamefont {H.}~\bibnamefont {Casini}}, \bibinfo {author} {\bibfnamefont {M.}~\bibnamefont {Huerta}},\ and\ \bibinfo {author} {\bibfnamefont {D.}~\bibnamefont {Pontello}},\ }\href {https://doi.org/10.1103/PhysRevD.98.125008} {\bibfield  {journal} {\bibinfo  {journal} {Phys. Rev. D}\ }\textbf {\bibinfo {volume} {98}},\ \bibinfo {pages} {125008} (\bibinfo {year} {2018})}\BibitemShut {NoStop}%
\bibitem [{\citenamefont {Pye}\ \emph {et~al.}()\citenamefont {Pye}, \citenamefont {Hingane},\ and\ \citenamefont {Jonsson}}]{longer_paper}%
  \BibitemOpen
  \bibfield  {author} {\bibinfo {author} {\bibfnamefont {J.}~\bibnamefont {Pye}}, \bibinfo {author} {\bibfnamefont {A.}~\bibnamefont {Hingane}},\ and\ \bibinfo {author} {\bibfnamefont {R.~H.}\ \bibnamefont {Jonsson}},\ }\href@noop {} {\bibinfo  {journal} {(in preparation)}\ }\BibitemShut {NoStop}%
\bibitem [{\citenamefont {Zhurina}\ and\ \citenamefont {Karmazina}(1966)}]{zhurina_tables_1966}%
  \BibitemOpen
\bibfield  {journal} {  }\bibfield  {author} {\bibinfo {author} {\bibfnamefont {M.~I.}\ \bibnamefont {Zhurina}}\ and\ \bibinfo {author} {\bibfnamefont {L.~N.}\ \bibnamefont {Karmazina}},\ }\href@noop {} {\emph {\bibinfo {title} {Tables and formulae for the spherical functions $P^m_{-1/2+i\tau}(z)$}}}\ (\bibinfo  {publisher} {Pergamon Press},\ \bibinfo {year} {1966})\BibitemShut {NoStop}%
\bibitem [{\citenamefont {Hobson}(1931)}]{hobson_theory_1931}%
  \BibitemOpen
  \bibfield  {author} {\bibinfo {author} {\bibfnamefont {E.~W.}\ \bibnamefont {Hobson}},\ }\href@noop {} {\emph {\bibinfo {title} {The theory of spherical and ellipsoidal harmonics}}}\ (\bibinfo  {publisher} {Cambridge University Press},\ \bibinfo {year} {1931})\BibitemShut {NoStop}%
\bibitem [{\citenamefont {Arias}\ \emph {et~al.}(2017)\citenamefont {Arias}, \citenamefont {Casini}, \citenamefont {Huerta},\ and\ \citenamefont {Pontello}}]{arias_anisotropic_2017}%
  \BibitemOpen
  \bibfield  {author} {\bibinfo {author} {\bibfnamefont {R.~E.}\ \bibnamefont {Arias}}, \bibinfo {author} {\bibfnamefont {H.}~\bibnamefont {Casini}}, \bibinfo {author} {\bibfnamefont {M.}~\bibnamefont {Huerta}},\ and\ \bibinfo {author} {\bibfnamefont {D.}~\bibnamefont {Pontello}},\ }\href {https://doi.org/10.1103/PhysRevD.96.105019} {\bibfield  {journal} {\bibinfo  {journal} {Phys. Rev. D}\ }\textbf {\bibinfo {volume} {96}},\ \bibinfo {pages} {105019} (\bibinfo {year} {2017})}\BibitemShut {NoStop}%
\bibitem [{\citenamefont {Wolf}(2008)}]{wolfNotSoNormalModeDecomposition2008}%
  \BibitemOpen
  \bibfield  {author} {\bibinfo {author} {\bibfnamefont {M.~M.}\ \bibnamefont {Wolf}},\ }\href {https://doi.org/10.1103/PhysRevLett.100.070505} {\bibfield  {journal} {\bibinfo  {journal} {Phys. Rev. Lett.}\ }\textbf {\bibinfo {volume} {100}},\ \bibinfo {pages} {070505} (\bibinfo {year} {2008})}\BibitemShut {NoStop}%
\end{thebibliography}%

\end{document}